\documentclass[twoside]{report}
\usepackage{iwsm}
\usepackage{graphicx}
\usepackage{amsmath, amssymb}
\usepackage{booktabs}

\begin{document}


\title{Statistical shape analysis in a Bayesian framework for shapes in two and three dimensions.}
\titlerunning{Statistical shape analysis}

\author{Thomai Tsiftsi \inst{1}}
\authorrunning{Tsiftsi}    

\institute{University of Bath, Department of Mathematical Sciences, Bath, UK. }

\email{t.tsiftsi@bath.ac.uk}

\abstract{In this paper, we describe a novel shape classification method which is embedded in the Bayesian
paradigm. We discuss the modelling and the arisen shape classification algorithm for two and three dimensional data shapes. We conclude by evaluating the efficiency and efficacy of the proposed algorithm on the Kimia shape database for the two dimensional case.}



\maketitle



\section{Introduction}
Shape is an important feature of objects; it can be used in many applications such as the recognition and classification of objects in images. In the approach we take we represent such objects and their boundaries as continuous planar curves (i.e. one-dimensional lines which denote the outline of the object) and study their shapes. Our goal is to develop shape models, statistical procedures and classification methods of continuous planar shape curves and establish the statistical framework needed for their classification. In particular, we study how to classify shapes that are generated by such curves and how we can probabilistically assign them into their respective categories; given a set of pre-determined classes we would like to classify the observed data shapes -- we here define a \textbf{data shape} $y$ to be one of the shapes that we observed i.e. an ordered set of points in $\mathbb{R}^{2}$ or $\mathbb{R}^{3}$. These questions occur in many applications of shape modelling and thus are of broad interest.

\section{Modelling and classification}
The problem of classification can be mathematically formulated as the posterior probability of the class in question given the observed data; that is by $\mathbb{P}(C|{\boldsymbol{y}})$ where $C \in \mathcal{C}$ the class of the object and $\boldsymbol{y}  \in Y $ the set of all the observed data shapes. In a Bayesian framework, classification is performed by maximising the posterior probability of the class which by Bayes' theorem is: $\mathbb{P}(C|{\boldsymbol{y}})\propto \mathbb{P}({\boldsymbol{y}}|C)\mathbb{P}(C)$. For simplicity we choose the prior $\mathbb{P}(C)$ over the classes to be uniform although it can be freely chosen. The major task is then to calculate the likelihood which we partition over nuisance parameters that correspond to the data formation process; this implies the marginalisation of the likelihood over similarity transformations, namely translations, scales and rotations $g \in G \equiv \mathbb{R}^m \ltimes (\mathbb{R}^+ \times SO(m))$, bijections $b :  [1, ...n] \rightarrow [1, ...n]$, shape curves $\beta \in \mathcal{B} \equiv R^{m \times n}/(R^{m} \times (R^+ \times SO(m)))$, sampling functions $s \in \mathcal{S}$   and the inherent observational noise $\sigma$. For our applications we choose the observational model to represent errors in shape point collection as additive Gaussian white noise so that the likelihood function for the
complete data is given by:

\begin{multline}
\mathbb{P}(\boldsymbol{y}| C) = \sum_{b \in \mathcal{B}} \int \mathcal{D}\beta~ \mathcal{D}s~ \mathcal{D}g~  d\sigma~   \mathbb{P}(b)\mathbb{P}(s)\mathbb{P}(g)\mathbb{P}(\sigma)\mathbb{P}(\beta|C) \\ \times 
\exp{\left(-\frac{1}{2\sigma^2}\sum_{i=1}^{n} |\boldsymbol{y_{{i}}}-g \circ \boldsymbol{\beta} (s(b_{i}^{-1}))|^2 \right)}
\label{tsiftsi:lhdmodel}
\end{multline}

with a number of simplifying independence assumptions made. In order to estimate the posterior probability of a class, one should evaluate the sums and integrals over the nuisance parameters. In the next sections we discuss our computational strategies for dealing with these evaluations for both two and three dimensional shape data.

\section{The two dimensional case}
Our main goal is to evaluate the integrals in expression (\ref{tsiftsi:lhdmodel}) and thus perform Maximum a Posteriori (MAP) of a certain class given the data. In previous work, \emph{e.g.} Dryden and Mardia (1998) and Srivastava and Jermyn (2009) the integrations over the nuisance parameters were evaluated numerically by a zeroth order Laplace approximation. In Tsiftsi et al. (2014) we introduced an analytic way of carrying out the group integrations and the integrations over $\sigma$ resulting in a closed form expression.

\smallskip
To achieve that, we had to make an appropriate and statistically significant choice of priors. Initially, we used Jeffreys' joint prior for $g \in G \equiv \mathbb{R}^2 \ltimes (\mathbb{R}^+ \times SO(2))$ and $\sigma$ however due to induced divergences a regularized version was employed. Although this broke the invariance of the original posterior, the result of this integration was found to be:

\begin{align}
\mathbb{P}({y}|b,\beta,s) & = \frac{1}{Z} ~ \sum_{b \in \mathcal{B}} \int~ \mathcal{D}\beta ~ \mathcal{D}s ~  \left[  \tilde{n}\widetilde{\text{Var}(\boldsymbol{y})} - \frac{\tilde{n}^2\left|\widetilde{\text{Cov}(\boldsymbol{v},\boldsymbol{y})}\right|^2}{\tilde{n}\widetilde{\text{Var}(\boldsymbol{v})}+1/B^2}+2\zeta\right]^{-n-\alpha} \nonumber \\ 
&~~~~~~~~~~~~~~~~~~~~~~~~~~~~~~~~~~~~~~~~~~~~~~~~~~~~~~~  \times \mathbb{P}(b)\mathbb{P}(s)\mathbb{P}(\beta|C)
\label{tsiftsi:complete}
\end{align}

where $B,\alpha, \zeta$ are appropriate regulators, $Z$ the normalisation constant, $n$ the  sample points and $\widetilde{\text{Cov}(\boldsymbol{v},\boldsymbol{y})}=\frac{1}{\tilde{n}}\left[ \sum_{i}{v_{i}} {\bar{y}_{i}}-\frac{1}{\tilde{n}}\sum_{i}\sum_{j}{v_{i}} {\bar{y}_{j}}\right]$. For details regarding the priors and the calculations refer to Tsiftsi et al. (2014). The proposed algorithm returns high classification rates. To demonstrate its efficiency we encounter an example on two shape databases in section \ref{tsiftsi:example}.

\section{The three dimensional case}
Another problem of interest is the generalisation of the previous case to its three dimensional equivalent by assuming that $y \in \mathbb{R}^3$. The three dimensional case is treated in a similar way
as the two dimensional case: our goal is to classify a shape by performing MAP on the class $C$. We follow the same steps as in the two-dimensional case and we marginalise the likelihood over the nuisance parameters that take part in the data formation process however similarity transformations are now represented by $g \in G \equiv \mathbb{R}^3 \ltimes (\mathbb{R}^+ \times SO(3))$ since $y \in \mathbb{R}^3$.

\smallskip
The challenge is the analytic evaluation of the integrals of the marginalised likelihood and especially the integration over three-dimensional rotations. Initially, translations were integrated against the uniform Haar measure. The result, as expected, was analogous to the two dimensional case and had to be integrated with respect to rotations. For the integration, we chose to represent rotations as unit quaternions; the full quaternionic space is described by: $\mathbb{H}=\{a + bi + cj + dk ∶ a, b, c, d \in R \}$ with $i, j, k$ the three special unit imaginary quaternions. An important point about quaternions, is the fact that they do not commute. The basis quaternions anti-commute and they provide a representation of $SU (2)$. The value of the commutator is $[y, q] = 2 \underline{\boldsymbol{y}} \times \underline{\boldsymbol{q}}$, with $y, q \in \mathbb{H}$ and $\underline{\boldsymbol{y}}, \underline{\boldsymbol{q}}$ their
vectorial parts.

\smallskip
For the integration over quaternions, we choose to integrate over the full quaternionic space $\mathbb{R}^4$ and impose a constraint that takes into account only unit quaternions. Since unit quaternions live on the surface of the unit 4-sphere, we impose the constraint $\delta(|q|^2 - 1)$. This $\delta$-function is invariant under the action of $SU(2)$ on the parameters since rotations act by isometries and thus do not change the length of the quaternions. Integrating the result of translations over rotations by imposing the constraint one has:

\begin{align}
P(y | b, \beta, s, \sigma) & = \int d^4q ~ \delta(|q|^2-1)~ \exp \left(     \frac{|\sum_{i}^{N}\boldsymbol{Y}_{i}|^2}{2n\sigma^2} - \frac{\sum_{i}^{N}|\boldsymbol{Y}_{i}|^2}{2\sigma^2}\right)
\label{tsiftsi:delta}
\end{align}

\noindent
To perform the integration, we replace the $\delta$-function by its Fourier equivalent which introduces a second integration that can be simplified to:

\begin{align}
\nonumber
P(y|b, \beta, s, \sigma) & = \iint dk ~ d^4q ~ \exp\left(ik(|q|^2-1)  \right) \exp \left(     \frac{|\sum_{i}^{N}\boldsymbol{Y}_{i}|^2}{2n\sigma^2} - \frac{\sum_{i}^{N}|\boldsymbol{Y}_{i}|^2}{2\sigma^2}\right) \\
& =  \frac{1}{2\pi}\iint dk~ d^4q ~ \exp\left(-ik \right) \exp \left(4n~ [q^{T}M(k)q] \right) 
\label{tsiftsi:deltaFourier}
\end{align}

\noindent
where $M_{ij}(k)=ik \delta_{ij}+\delta_{0i}(\overline{\underline{\boldsymbol{\hat{v}}}^{T}\times \underline{\boldsymbol{\hat{y}}}})_{i}+(1-\delta_{01})(1-\delta_{0i})\left[\overline{\underline{\boldsymbol{\hat{y}}} \otimes \underline{\boldsymbol{\hat{v}}}})_{ij} - \delta_{ij}\overline{\underline{\boldsymbol{\hat{v}}}^{T}\times \underline{\boldsymbol{\hat{y}}}})\right]$ the symmetrised, positive definite $4 \times 4$ covariance matrix of the $q$ components.

\smallskip
We now discuss the integral over the quaternionic parameters that generate the $SO(3)$ rotations. We followed Wood (1993) and calculated the appropriate Haar measure for the quaternionic representation which was proven to be related to the normalisation constant of the Bingham distribution. The result of integrating over $k$ will supply the Haar measure on the space of unit quaternions and restrict our parameters $q$ to this surface. By diagonalising $M$ we can rewrite the exponent of expression (\ref{tsiftsi:deltaFourier}) as:

\begin{align}
\exp \left(4n~[q^{T}M(k)q] \right) = \int_{S^3}  \exp\left(4n ~ \sum_{i}\lambda_{i}\tilde{q}^2_{i}~d[\tilde{q}] \right)
\end{align}

Here, the $\tilde{q}_{i}$ generate rotations in $SO(3)$ which will be uniformly distributed \textbf{if and only if} the $\tilde{q}_{i}$ are uniform on a unit hemisphere in $\mathbb{R}^4$. This means that choosing the usual uniform measure on $S^3$ for $d[\tilde{q}]$ induces the Haar measure on the space of rotations. This ensures that the chosen measure in (\ref{tsiftsi:deltaFourier}) is the appropriate one and induces invariance under the act of rotations so that we do not favour one rotation over another.

\smallskip
Returning to expression (\ref{tsiftsi:deltaFourier}), it is easy to see that the integral with respect to $q$ refers to a multivariate Gaussian distribution. Assuming that the eigenvalues of matrix $M$ are negative the evaluation of the quaternionic integral of this multivariate Gaussian distribution is:

\begin{align}
\nonumber
P(y|b, \beta, s, \sigma) & =  \frac{1}{2\pi}\iint d^4q~ dk~  \exp\left(-ik \right) \exp \left(4n~ [q^{T}M(k)q] \right) \\
& \propto \frac{1}{2\pi}\int dk ~ \exp\left(-ik \right) \frac{4~n~\pi^2}{\sqrt{\text{det(M)}}}
\label{tsiftsi:kintegral}
\end{align}

with $\text{det(M)}$ the determinant of matrix $M$ which has $k$ dependence and is invariant to rotations of  $y$ since it has been written in a manifestly rotationally invariant way. It is common practice to evaluate integrals of this form by contour integration. For this we would have to promote k to the complex plane and choose an appropriate path in the $k$-plane. Since the expression of the determinant is not a perfect square, the presence of the square root in the denominator  instead of turning points at which the denominator vanishes into poles, it turns them into branch cuts making the integration over these extremely difficult. Thus contour integration cannot be of help and the integral over $k$ cannot be done analytically.

\smallskip
We were thus forced to Taylor expand the square root in the denominator of expression (\ref{tsiftsi:kintegral}) in order to be able to analytically approximate the integral. However, this represents an important step towards generalising our work on planar shapes to three-dimensional curves. The calculations of the integration of the remaining nuisance parameters are challenging, although positive developments have been made towards a series solution. We leave the remaining calculation for
future consideration as an extension of the analysis presented here. This work is still in progress but shows promising signs of improving upon the current shape classification methods in three-dimensions.

\section{Example in two-dimensions}\label{tsiftsi:example}
In order for the algorithm's efficacy on the classification of two-dimensional data shapes to be tested and verified, examples from two shape databases were considered: the Kimia and a simulated letter database. In the latter case the application of our algorithm comes with a warning; ordinarily the orientation of letters is crucial (for example W versus M and C versus U) whereas our likelihood has been constructed to be invariant under rotations of the data. The
tests on this database should be understood as a general test of our algorithm which is used for demonstrational purposes and not as a serious proposal for recognition of written letters.

\begin{figure}[!ht]\centering
\includegraphics[width=0.7\textwidth]{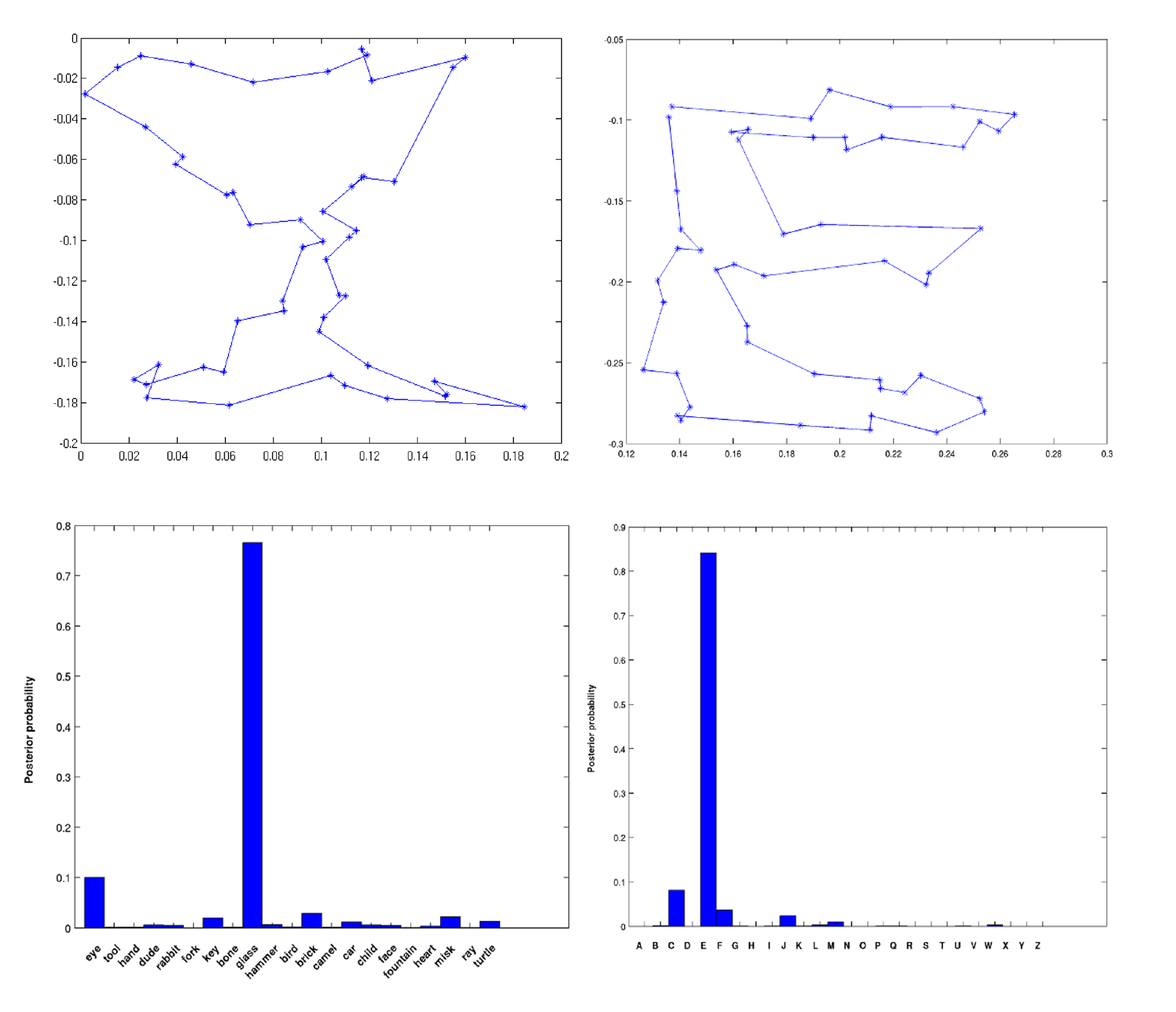}
\textbf{\caption{\label{tsiftsi:glass}  Classification results for a Kimia shape and the letter E.}}
\end{figure}

\smallskip
 Both databases were comprised of binary images which were used for training and testing purposes. The shapes' boundaries were extracted by MATLAB built-in functions and simulated shapes played the role of the observed data sets. The proposed algorithm was tested on the simulated data sets and its classification results are very positive, as is illustrated in Figure \ref{tsiftsi:glass}.

\smallskip
For the Kimia database we found that for 10 runs of 10 shapes each, the average classification level was $\hat{\mu} = 59\% \pm 7\%$ with the average success rate being more than $\hat{\mu}=80\% \pm 5\%$. From these experiments we concluded that the number of sampled points is crucial since as soon as the number of points increases to more than 50 the confidence levels become almost 90 percent. For the alphabet database,
the results for the average classification level were $\hat{\mu}= 77\% \pm 5\%$ with the average success rate $\hat{\mu}= 73\% \pm 6\%$. The evaluation of the performance of the algorithm in three-dimensions could be tested by using examples from 3D geological sand formations as previously discussed in Tsiftsi et al (2014).

%
%
%
\references
\begin{description}
\item[Dryden, I.L. and Mardia, K] (1998).
    {\it Statistical shape analysis}.
    J. Wiley.

\item[Kimia, B.B] (2015),
  Kimia database.
 {\it Available at www.lems.brown.edu/~dmc/}.

\item[Srivastava, A. and Jermyn, I.H.] (2009), Looking for shapes in 2D cluttered point clouds, {\it IEEE Trans. Patt.\ Anal.\ Mach.\ Intell.}, {\bf 31(9)},
    1616\,--\,1629.

\item[Tsiftsi, T. , Jermyn, I. and Einbeck, J.] (2014) Bayesian shape modelling of cross-sectional geological data, {\it in 29th International Workshop on Statistical Modelling, 14-18 July 2014, Goettingen, Germany; proceedings}, Amsterdam: Statistical Modelling Society, 161\,--\,164, \newline
[arXiv:1802.09631 [stat.ME]].

\item[Wood, A.T.A.] (1993), Estimation of the concentration parameters of the Fisher matrix distribution on ${SO}(3)$ and the Bingham distribution on ${S}_{q}, q \geqslant 2$,
{\it Australian Journal of Statistics},
{\bf 35 (1)}, 69\,--\,79.

\end{description}

\end{document}